\documentclass[superscriptaddress,twocolumn,showpacs,preprintnumbers,amsmath,amssymb,prb]{revtex4-1}
\usepackage{graphicx,times,epsfig,color}

\usepackage{graphicx,times,color}
\usepackage{psfrag}
\usepackage{epsfig}
\usepackage{verbatim}
\usepackage{graphicx}
\usepackage{amsmath}
\usepackage{amsfonts}
\usepackage{amssymb}
\usepackage{ulem}

\begin{document}

\title{Transport regimes in a double quantum dot device}

\author{L. Costa Ribeiro}

 \affiliation{Centro Federal de Educa\c{c}\~ao Tecnol\'{o}gica Celso Suckow da Fonseca (CEFET-RJ/UnED-NI), RJ 26041-271, Brazil.}

 \affiliation{Departmento de F\'{\i}sica Aplicada, Unidad Asociada del Consejo Superior de Investigaciones Cient\'{\i}ficas and Instituto Universitario de Materiales, Universidad de Alicante, San Vicente del Raspeig, Alicante 03690, Spain.}

\author{I. J. Hamad}
\affiliation{Departamento de  F\'{\i}sica, Pontificia Universidade Cat\'olica do Rio de Janeiro (PUC-Rio), 22452-970, Caixa Postal: 38071 Rio de Janeiro, Brazil.}
\affiliation{Instituto de F\'isica Rosario. Universidad Nacional de Rosario. Bv. 27 de Febrero 210 bis, Rosario (2000), Argentina.}

\author{G. Chiappe}
 \affiliation{Departmento de F\'{\i}sica Aplicada, Unidad Asociada del Consejo Superior de Investigaciones Cient\'{\i}ficas and Instituto Universitario de Materiales, Universidad de Alicante, San Vicente del Raspeig, Alicante 03690, Spain.}

\author{E. V. Anda}
\affiliation{Departamento de  F\'{\i}sica, Pontificia Universidade Cat\'olica do Rio de Janeiro (PUC-Rio), 22452-970, Caixa Postal: 38071 Rio de Janeiro, Brazil.}

\date{\today}

\begin{abstract}

We analyze the transport properties of a double quantum dot device with both dots coupled to perfect conducting leads and to a finite chain of N non-interacting sites connecting both of them. The inter-dot chain strongly influences the transport across the system and the Local Density of States of the dots. We study the case of small number of sites, so that Kondo box effects are present, varying the coupling between the dots and the chain. For odd N and small coupling  between the inter-dot chain and the dots, a state with two coexisting Kondo regimes develops: the bulk Kondo due to the quantum dots connected to leads and the one produced by the screening of the quantum dots spins by the spin in the finite chain at the Fermi level. As the coupling to the inter-dot chain increases, there is a crossover to a molecular Kondo effect, due to the screening  of the molecule (formed by the finite chain and the quantum dots) spin by the leads. For even N the two-Kondo temperatures regime does not develop and the physics is dominated by the usual competition between Kondo and antiferromagnetism between the quantum dots. We finally study how the transport properties are affected as N is increased. For the study we used exact multi-configurational Lanczos calculations and finite $U$ slave-boson mean-field theory at $T=0$. The results obtained with both methods describe qualitatively and also quantitatively the same physics.

\end{abstract}
\pacs{73.63.Fg, 71.15.Mb}
\maketitle

\section{Introduction}

Since the prediction of the occurrence of the Kondo effect \cite{Ko64} in a single quantum dot (QD) device \cite{GlRa88,NgLe88} and its subsequent experimental observation \cite{GoSh98}, several single, double (DQD) and multiple quantum dots devices, or systems with atoms or molecules acting as magnetic impurities, have been studied both theoretically \cite{GeMe99,Busser00,Davidovich02,Ribeiro12,Hamad13} and experimentally \cite{CheCha04,WiFr03,ChoJa12,NeBe11,Bork11}. The interest in these systems stems from their potential applications to quantum and classical computing \cite{LoDi98,DiLy04} and their usefulness as model systems to study the physics of strongly correlated electrons.

In a QD connected to leads the charge can be manipulated by means of an external gate potential. When the charge on the QD is close to an odd integer, the Kondo effect takes place and it results in perfect transmission through the system at temperatures below the Kondo temperature $T_K$ and for gate voltages that do not incorporate extra charge into the QD. The electrons in the QD and leads form a spin singlet, which is one of the most clear benchmarks of the Kondo effect \cite{He93}. The system constitutes an experimental realization of the single impurity Anderson model \cite{An61}.

Likewise, two QDs directly coupled between them amount to an experimental realization of the two-impurity Anderson model \cite{AlAn64}. Here the physics is much richer, particularly in the regime where each dot contains an unpaired electron. In this case, the state of the system is characterized by the competition between the tendency of the conduction electrons on the leads to screen the spins localized in the QDs and the antiferromagnetic (AF) coupling between these spins \cite{GeMe99}. The former favors the formation of a two Kondo singlet, while the latter corresponds to a molecular state constituted by the two QDs. The resulting ground state and transport properties of the system depend sensitively on the relative strength of the interactions and the topology of the system.  In each scenario, phase transitions are predicted to occur between different quantum states. These states can be characterized by a single Kondo resonance, or by a Kondo peak with a very narrow dip at the Fermi level which is representative of a two stage Kondo regime \cite {CoGr05,Anda08}, or even by a situation in which there is a splitted resonance resultant of the dominance of the AF correlations between QDs. In this latter case the conductance changes from a large value to zero, in a wide region of the applied gate voltage. Moreover, it has been predicted that this system could suffer a quantum phase transition, which involves a non-Fermi liquid fixed point \cite{JaKr81}, which in fact has proven extremely difficult to be observed \cite{Malecki10,Jayatilaka11,Bork11}. These transitions in the two-impurity Anderson model have received wide attention in the theoretical and experimental literature in recent years \cite{GeMe99,ChoJa12}. An understanding of the physics of two interacting impurities is important and believed to play a role in the electronic properties of a wide range of strongly correlated materials, including spin glasses and heavy fermion compounds \cite{He93}.

It is known that when an impurity is coupled to a finite number of N non-interacting sites, the Kondo effect modifies its character in what is called the ``Kondo Box'' effect, a phenomenon that has also been theoretically \cite{Thimm99,Simon02_03,Cornaglia02,Schlottmann01} and experimentally \cite{Ralph95} studied. In this case, the finite system's density of states (DOS) consists of a series of peaks separated by an energy $\Delta$ inversely proportional to $v/N$ where $v$ is the hopping matrix element among the sites and $N$ is the number of them. The peaks could have a finite width $\delta$ if the quantum box is weakly connected to an infinite system. When N is very large these peaks overlap and the DOS resembles that of a continuum. As N is decreased or $v$ augmented, finite size effects start taking place when $\Delta \approx T_K$ \cite{Thimm99}, if $\delta<<\Delta$. Depending on whether one of these peaks coincide with the Fermi level or not ('at resonance' or 'off resonance' case), there is a single or a splitted resonance in the impurity's Local DOS (LDOS) \cite{Thimm99}, and the Kondo temperature behaves in a different way \cite{Simon02_03}.

The physics of a Kondo Box described above can be realized in systems of two impurities coupled between them by a finite number of non-interacting sites. This problem is thus very interesting and has also been theoretically studied  in various configurations \cite{Simon05, Galkin04, Durganandini06}. Here, finite size effects can take place together with a magnetic interaction between the impurities of the Ruderman-Kittel-Kasuya-Yosida (RKKY) type \cite{Rudermann54}. In this case, when the Kondo effect is present, the fouth-order RKKY interaction between the impurities is mediated by the electrons of the non-interacting sites, that are participating simultaneously in the Kondo screening of each impurity, as was studied in Ref. \cite{Simon05}. There has also been pointed out that other types of magnetic interactions arise such as the 'Kondo correlated' or supeerexchange interactions studied in Ref. \cite{Ong11}. Also, using variational wavefuntions, it was predicted  \cite{Simonin06} that the interaction between the impurities is mainly due to an interference-enhanced hybridization that generates Kondo doublet states. This interaction can be more important than the RKKY type and of ferromagnetic character. 

Experimentally, these kind of configurations have also been studied, as was done recently \cite{Neel11} with a system of two Co impurities interacting throught N Cu atoms placed between them, constituting a linear $CoCu_NCo$ cluster deposited on a surface of Cu(1,1,1). It was observed that the number N of Cu in the linear chain strongly influences the Kondo temperature $T_K$ of each impurity,  in general lowering it and having an oscillatory behavior for $N>3$ compatible with RKKY interactions \cite{Neel11}.

In this paper, we analyze a DQD (or double impurity) system simultaneously connected to metalic leads and between themselves through a finite chain. The study emphasizes the fact that, as mentioned above, the impurities in such a system are coupled to, and interact through, a non interacting linear chain (NILC) that constitutes a quantum box, whose electrons can participate in the Kondo screening. This implies that the physics corresponding to this problem is one in which there is a interplay between a bulk continuous Kondo regime and a two impurity Kondo box. We analyze the transport properties for different values of the number N of sites of the NILC and for different couplings of the QDs with it. We study the system at $T=0$, but the results are valid also for temperatures well below the characteristic single impurity Kondo temperature. In particular, we present a detailed analysis of NILC with $N=1$ and $N=2$, that are limiting cases of a more general situation in which discrete systems present resonances at the Fermi level or not, respectively.  We ask ourselves what is the effect of varying the connection of the impurities to the finite system. In doing so, we are able to characterize a crossover, that occurs for N=1 (odd N in general), from a two Kondo temperature regime to a molecular Kondo regime, in which three resonances rise up as the coupling between the NILC and QDs is increased. This is reflected in the LDOS and in the transmission of the system.  For even N, a crossover from a single central peak to a splitted peak is observed in the LDOS, in the transmission and in the conductance. In this case, the system behaves in a similar way as one with a direct connection between the impurities \cite{Ribeiro12,Hamad13,Busser00}. Finally, we also present results for larger N, but small enough so as to guarantee that finite-size effects are still predominant. We conclude that the characteristics of the LDOS and the conductance (or transmission) are largely influenced by the spectrum of the quantum box formed by the two dots plus the NILC.

The paper is organized as follows. In section \ref{ModelandHamiltonian} we present the model and the Hamiltonian we used to study it. In Section \ref{Methods} we briefly describe the methods used to study the system, while in section \ref{Results} we present the results. Finally, we make the concluding remarks in \ref{Conclusions}.

\section{Model and Hamiltonian}
\label{ModelandHamiltonian}

\begin{figure}
\centering 
\rotatebox{0}{\scalebox{0.33}{\includegraphics{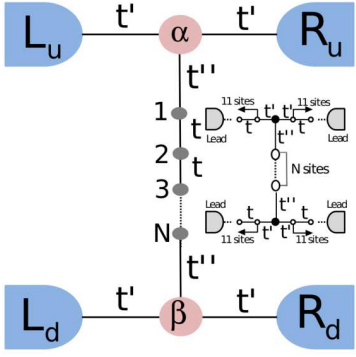}}}
\caption{(Color online). The structure studied in this work. Inset: The scheme used for the Multi Configurational Lanczos calculations (see section \ref{Lanczos})}
\label{figura1}
\end{figure}

The two QDs system is sketched in Fig.[\ref{figura1}]. It is described by an Anderson Hamiltonian composed by three terms, $H=H_{0}+H_{t}+H_{lcc}$. The first contribution carries the local physical information of the QDs and is given by,
\begin{eqnarray}
H_{0}&=&\sum_{i={\alpha;\beta}\atop{\sigma}}{\epsilon}_{i}f^{\dagger}_{i\sigma}f_{i\sigma}+
\sum_{i={\alpha;\beta}}U f^{\dag}_{i\downarrow}f_{i\downarrow}f^{\dag}_{i\uparrow}f_{i\uparrow},
\end{eqnarray}

\noindent where $\epsilon_{i}$, $U$ and $f^{\dag}_{i\sigma}(f_{i\sigma})$ represent, respectively, the local energy state, the electron-electron Coulomb interaction and the operator that creates (annihilates) an electron with spin $\sigma$ in the i-th QD. We assume $\epsilon_i=0$. The local energy of the dots is tunned by a gate potential $V_{g}$ that, for simplicity, is considered to be the same for both  QDs. The second term can be written as,
\begin{eqnarray}
H_{t}=t'\sum^{\sigma}_{i=L;R \atop{j=\alpha;\beta} }(c^{\dag}_{i\sigma}f_{j\sigma}+H.c)+t''\sum_{{i=1;N} \atop{ j=\alpha;\beta \atop \sigma}}(f^{\dag}_{j\sigma}c_{i\sigma}+H.c)
\end{eqnarray}
and describes the connections of the QDs to both, the metallic leads and the NILC through the hopping terms $t'$ and $t''$, respectively. As the physics depends upon the relative values between these two parameters plus the Coulomb repulsion $U$, we fix $t'$ and $U$ and concentrate on the effects of varying $t''$. The third therm,
\begin{eqnarray}
H_{lcc}&=&2t\sum_{\sigma} (c^{\dag}_{i\sigma}c_{i+1,\sigma}+c^{\dag}_{-i\sigma}c_{-i-1,\sigma}+H.c)+\nonumber\\
&&+t\sum_{\sigma} (c^{\dag}_{i\sigma}c_{i+1,\sigma}+H.c)
\end{eqnarray}
describes the leads and the NILC.

The Fermi level in the reservoirs is adjusted to zero, $E_{f}=0$, while the hopping $t$ is adopted as the energy unit.

\section{Methods}
\label{Methods}

\subsection{Finite-U Slave Boson Mean Field Approximation.}

The main idea concerning the Slave Boson approach \cite{Coleman84,Kotliar86} consists in enlarging the Hilbert space by introducing in the Hamiltonian a set of bosonic operators which incorporate into the system the physics underlying the Kondo regime. These operators, $e_{i}$, $p_{i\sigma}$ and $d_{i}$, with $i$ corresponding to the i-th impurity, are responsible for projecting the system on a state of zero, single and double occupation on the impurity and are introduced by the hybridization of the Fermion operator which creates (annihilates) an electron with spin $\sigma$ in the i-th impurity,
\begin{eqnarray}
f^{\dag}_{i\sigma} \rightarrow Z_{i\sigma}^{\dag}f^{\dag}_{i\sigma},
\end{eqnarray}
where
\begin{eqnarray}
\label{z}
Z_{i\sigma}^{\dag}&=&[1-d^{\dag}_{i}d_{i}-p^{\dag}_{i\sigma}p_{i\sigma}]^{-\frac{1}{2}}(e^{\dag}p_{i\sigma}+
p^{\dag}_{i\bar{\sigma}}d_{i})\nonumber\\
&&\times[1-e^{\dag}_{i}e_{i}-p^{\dag}_{i\bar{\sigma}}p_{i\bar{\sigma}}]^{-\frac{1}{2}}
\end{eqnarray}

\noindent is an operator that, in the mean field approximation, becomes a parameter $\bar{Z}$ that reproduces correctly the results expected in the non-interacting limit $U=0$ \cite{Kotliar86} and is responsible for the renormalization of the connections of the impurities to the NILC and to the metallic leads. It is observed that this parameter becomes less than one, $\bar{Z}<1$, as the system enters into the Kondo regime.

The term "Slave Boson" comes from the constraints,
\begin{eqnarray}
e^{\dag}_{i}e_{i}+\sum_{\sigma}p^{\dag}_{i\sigma}p_{i\sigma}+d^{\dag}_{i}d_{i}-1=0
\end{eqnarray}
and
\begin{eqnarray}
f^{\dag}_{i\sigma}f_{i\sigma}-p^{\dag}_{i\sigma}p_{i\sigma}-d^{\dag}_{i}d_{i}=0,
\end{eqnarray}

\noindent that are imposed to the bosons operators in order to eliminate the non-physical states, assuring that the impurity is occupied with zero, one or two electrons and establishing a correspondence between bosons and fermions. They are incorporated to the Hamiltonian through the Lagrange multipliers $\lambda^{i}_{1}$ and $\lambda^{i}_{2\sigma}$.

Within the slave boson mean field approximation at finite U (finite-U SBMFA) we write the effective Hamiltonian,
\begin{eqnarray}
H_{eff}&=&\sum_{i=\alpha,\beta\atop{\sigma}}\epsilon_{i}f^{\dagger}_{i\sigma}f_{i\sigma}+
t''\sum_{j=\alpha;\beta\atop{i=1;N\atop{\sigma}}}\bar{Z}_{j}(f^{\dag}_{j\sigma}c_{i\sigma}+H.c.)+\nonumber\\
&+&\sum_{i=\alpha,\beta}U_{i}{\langle}d_{i}{\rangle}^{2}+
t'\sum_{i=L;R\atop{j=\alpha;\beta\atop{\sigma}}}\bar{Z}_{j}(c^{\dag}_{i\sigma}f_{j\sigma}+H.c.)+\nonumber\\
&+&\sum_{i=\alpha,\beta}\lambda^{i}_{1}({\langle}e_{i}{\rangle}^{2}+{\langle}p_{i\sigma}{\rangle}^{2}+{\langle}d_{i}{\rangle}^{2}-1)+\nonumber\\
&+&\sum_{i=\alpha,\beta\atop{\sigma}}\lambda^{i}_{2\sigma}(f^{\dagger}_{i\sigma}f_{i\sigma}-{\langle}p_{i\sigma}{\rangle}^{2}-{\langle}d_{i}{\rangle}^{2})+H_{lcc},
\end{eqnarray}

\noindent where we observe that the local energy levels $\epsilon_{i}(i=\alpha;\beta)$ are renormalized by the Lagrange multiplier $\lambda^{i}_{2\sigma}(i=\alpha;\beta)$, $\epsilon_{i}+\lambda^{i}_{2\sigma}$ and the connections $t'$ and $t''$ of the QDs by the multiplicative slave boson parameter $\bar{Z}$. These are the two renormalizations that, in the context of the finite-U SBMFA, carry the system into the Kondo regime. At $T=0$ the effective Hamiltonian  $H_{eff}$ is minimized with respect to the Lagrange multipliers and to the mean values of the bosons operators, resulting in a non-linear set of ten equations that have to be solved self-consistently in order to obtain the numerical values of these quantities.

\subsection{Multi Configurational Lanczos}
\label{Lanczos}

In the Multi Configurational Lanczos (MCL) calculation \cite{Chiappe13} the system is described by the same tight binding based Hamiltonian discussed before. The process requires the definition of a sub-system, a cluster of an arbitrary number of M sites that includes the DQD, the NILC and a portion of the connecting leads. Due to numerical reasons, each part of the four leads is taken to be constituted by eleven non-interacting sites, as is shown in the inset of Figure \ref{figura1}. As N is the number of sites of the NILC, the size of the cluster taken is given by $M = N+2+44$. In each lead we do a transformation from left/right sites to symmetric/anti-symmetric channels. Only symmetric channels are coupled with the central structure (DQD + NILC).

In order to numerically obtain the ground state of the cluster we use a mean field solution as starting state. The Hilbert space is increased by successive applications of H. In each step we obtain the exact solution within the restricted Hilbert subspace considered, until the desired convergence is reached and the ground state is obtained. After this process, we proceed to calculate the Green functions of the finite cluster, using a Lanczos procedure.  Finally, the cluster is embedded into the rest of the system, within the LDECA formalism \cite{Anda08}, using a Dyson equation to obtain the Green function of the entire system.

\section{Results}
\label{Results}

In this section we study the transport properties of the system. To this purpose we calculate, using both methods described above, the local density of states (LDOS) at the QDs, the transmission $T$, as a function of $\omega$, and the conductance $G$ as a function of the gate potential $V_g$, or, equivalently, the energy levels of the impurities. We also present finite-U SBMFA results for the parameters $\bar{Z}$ and $\tilde{\epsilon}$, that help to understand the physics involved.

Conductance or transmission, magnitudes that can be experimentally measured, are calculated from the upper left to the upper right lead. The results were obtained for the system in different regions of the parameter space and are divided in $N=1$, $N=2$ and large N limit. The Coulomb interaction in the QDs and the connection with the leads are always $U=0.5$ and $t'=0.2$, respectively, in units of $t$.

\subsection{N=1}

\begin{figure}
\centering 
\rotatebox{0}{\scalebox{0.45}{\includegraphics{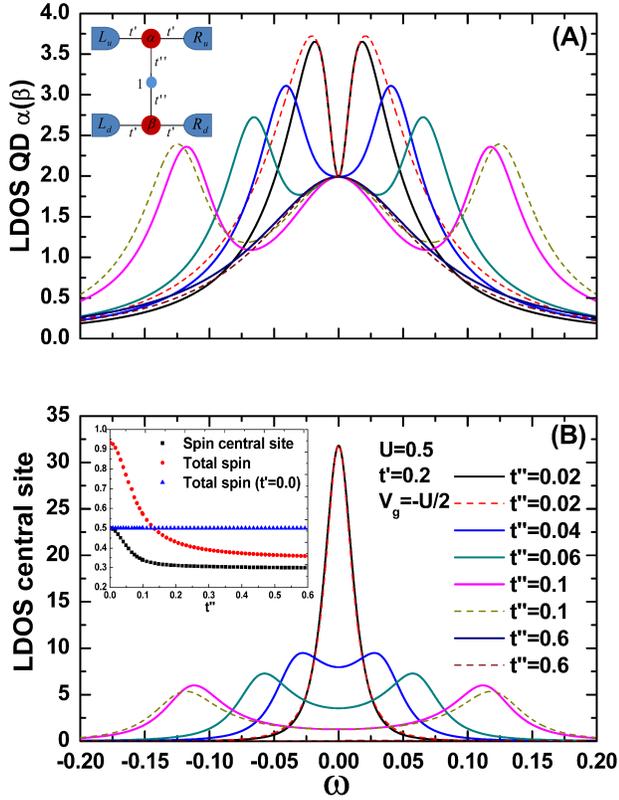}}}
\caption{(Color online) LDOS calculated (A) in the QDs and (B) in the non-interacting site 1 as a function of the energy $\omega$ for the N=1 case, and for different magnitudes of the connection $t''$ between the dots and the non-interacting central site at $V_g=-U/2$. The continuous lines corresponds to the finite-U SBMFA results, while the dashed lines correspond to MCL results. In the inset of panel (B) are shown the spin in the non-interacting central site (black squares), the total spin (red circles), and the total spin of the isolated molecule (blue triangles). See discussion in the text.}
\label{figura2}
\end{figure}

We take the $N=1$ case as representative of a quantum box with states at the Fermi level. For this case, it is instructive to think the two dots system plus the central non-interacting site as a three atoms molecule, with three molecular energy levels. One level has zero energy and weight only in the QDs. The other two levels have energy $\pm \sqrt{2} t'' \bar{Z}$ (where $\bar{Z}$ is the finite-U SBMFA renormalization parameter that, when squared, represents the weight of the Kondo state), whose larger weight falls on the central site. These levels exist since each dot is in a Kondo state due to its coupling to the respective leads, and hence has a resonance at zero energy. Therefore, the charge of the molecule is three electrons. In Fig. \ref {figura2} we present the LDOS calculated (A) in the QDs and (B) in the non-interacting central site, obtained with the gate potential $V_{g}$ adjusted in the particle-hole symmetric position $V_{g}=-U/2$. We can clearly identify two quantum regimes in the system. The first, for $t'' \lesssim 0.04$, in which we observe a peak with a narrow dip just at the Fermi level, is characterized by the existence of two energy scales that correspond to two Kondo temperatures, $T_{K1}$ and $T_{K2}$. These energy scales naturally emerge from the figure and are associated, respectively, with the widths of the dip and of the peak in the LDOS of the QDs. In this regime the particles, and hence the spins, are equally distributed in the three quasi-degenerated molecular energy states that exist next to $E_f=0$. Then, the first Kondo temperature can be thought of as coming from the partial screening of the QDs spin by the free spin allocated at the central non-interacting site. The remanent spin is then completely screened by the leads, characterizing the second Kondo temperature. 

\begin{figure}
\centering 
\rotatebox{0}{\scalebox{0.35}{\includegraphics{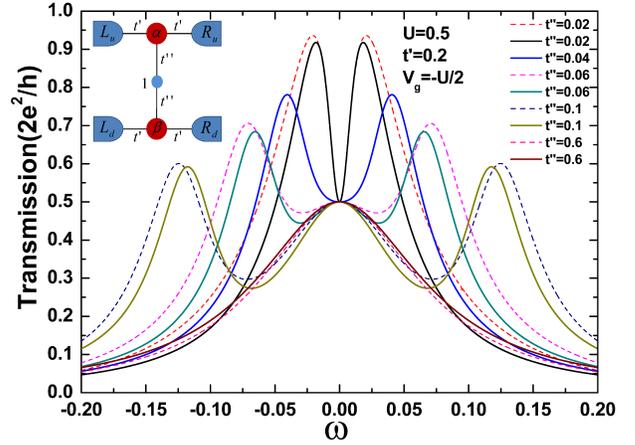}}}
\caption{(Color online) Transmission $T$ as a function of $\omega$ for the system with $N=1$, $V_{g}=-U/2$ and different magnitudes of the connection $t''$ with the central non-interacting site. The continuous lines correspond the finite-U SBMFA results, while the dashed lines correspond to MCL results.} 
\label{figura3}
\end{figure}

\begin{figure}
\centering 
\rotatebox{0}{\scalebox{0.48}{\includegraphics{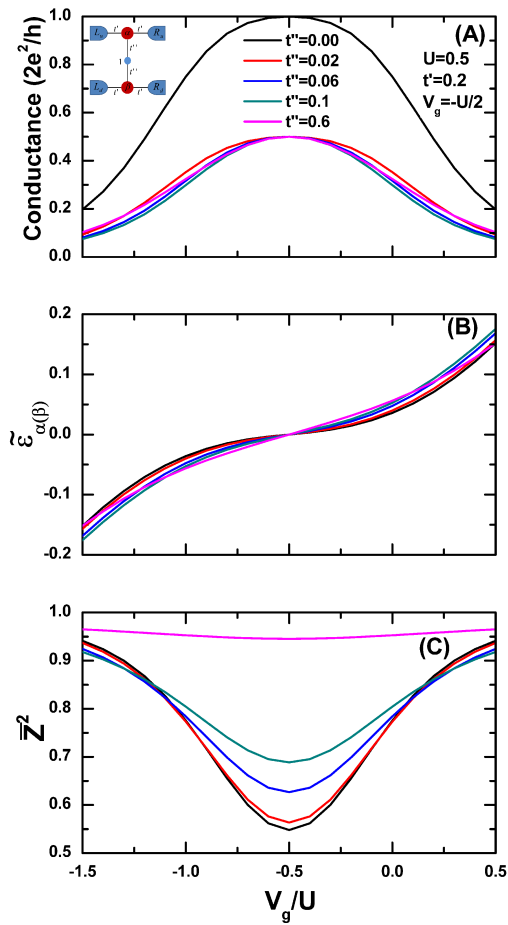}}}
\caption{(Color online) Conductance through the upper leads (A), renormalized energy level $\tilde{\epsilon}_{\alpha(\beta)}$ (B) and the parameter $\bar{Z}^2$ (C), as a function of the gate potential $V_{g}$ applied in the basis of the QDs for the system with different magnitudes of the connection $t''$ with the central non-interacting site.} 
\label{figura4}
\end{figure}

One of the key points to obtain this double Kondo temperature behavior is that the NILC shows, for low values of $t''$, a resonance in the LDOS just at the Fermi level. Then, there is a free spin allocated in this level that screens the spins at the QDs. The width of this resonance is similar to the width of the dip in the LDOS at the QDs. As already explained, when the molecule is disconnected from leads it has three levels: one at zero energy and weights only on the QDs and another two at energies $\pm \sqrt (2) t'' \bar{Z}$ with most of its weight on the central site. When connected, the peaks are broadened by their hybridization with the leads. While $t''$ is small enough as compared with the connection between the dots and the leads ($t'$), these two levels superpose creating a single resonance at the central site. The analysis of the central site's Green function shows that this occurs while $t'' \lessapprox T_K/(2 \sqrt(2\bar{Z})$, where we have assumed that the Kondo peaks have a width of $T_K$. For the parameters used in Fig. \ref{figura2}, this occurs when $t'' \lessapprox 0.02$, above which a crossover begins to a molecular Kondo regime, discussed below, and a two peaks splitted structure appears at the LDOS of the central site.  It is interesting to observe that the central resonance at the central site opens a tunneling channel, at the Fermi level, between the upper left and the down leads. As a consequence, the transmission at the Fermi level between the upper left and upper right leads, presented in Fig. \ref{figura3}, reaches the value $e^{2}/h$ which corresponds to half of its maximum $2e^{2}/h$, obtained when the transmission is analyzed displaced by a small amount outside the Fermi level. A similar result can be obtained by studying the behavior of the conductance shown in Fig.\ref{figura4} : conductance through the up (down) wire is exactly half of the value of that obtained when the dots are disconnected ($t'' = 0$). When the connection between the dots is established the electronic flow can be equally divided between the inter-dot channel and the upper left-right one, independently of the value of $ t''$ \cite{note1}.   

It is also important to notice that in this regime, the magnetic moment of the non-interacting site assumes its maximum value and as a consequence is capable of screening the dot's spins (see inset of Fig. \ref{figura2} B). To reinforce this image, note also that while the total spin for the free molecule is $S=0.5$, it changes to $S=1$ as it is connected to the leads, reducing its value as $t''$ increases. This $S=1$ value for $t''\approx 0$ is in fact a mean value between $S=0.5$ and $S=1.5$, the two possible values of the total spin of three independent spins. It shows that, as it is natural for $t'' \approx 0$, the molecule is not formed as a whole, something that happens when $t''$ increases, but all the same  connecting the molecule to the leads changes the internal spin correlations. This is a quite general fact also observed in other systems, as was shown for example in Ref. \cite{Strozecka12}, where the spin of MnPc changes from $1.5$ to $1$ by depositing it on a Bi(110) surface.

To emphasize the points mentioned above we have verified with MCL that, in the two Kondo temperatures regime, when there is a peak with a narrow dip in the LDOS of the dots, a magnetic field applied at the central non-interacting site produces the disappearing of the dip and the single Kondo peak is recovered. It is clear that for the central site to be able to participate in the screening of the QDs spins, its electron has to be at the Fermi level and with a fluctuating spin. The external magnetic field freezes the spin by opening a spin dependent Zeeman splitting at the central site, eliminating its screening capabilities. The recovering of the central Kondo peak as the central dip dissapears is a confirmation of this process and of the role played by the inter-dot site.

Note that our results differ from a study in a similar system in the so called ``at resonance'' situation \cite{Galkin04}. The two-Kondo temperature regime is manifested as a peak with a dip in the LDOS, while in the mentioned study there is only one Kondo resonance that results from the sum of two peaks with different widths, each one reflecting its corresponding Kondo temperatures. In our effective one body finite-U SBMFA Hamiltonian, after combining the upper and lower channels with a symmetric-antisymmetric transformation, a further symmetric-antisymmetric transformation with respect to the central site results in a symmetric pseudodot that is coupled to a reservoir \textit{and} to the single non-interacting central site, plus another independent antisymmetric pseudodot that is only coupled to the reservoir. The LDOS of the symmetric channel is a peak with a dip at the Fermi level, while the LDOS of the antisymmetric one is just a Lorentzian peak. The LDOS at the QDs can be written as the sum of the LDOS corresponding to each pseudodot, which gives the results of Fig. \ref{figura2}, that, we emphasize, have a very good agreement with the MCL results, where no approximation is made in the Hamiltonian. Besides, as $t''$ increases, there is a transition to a molecular Kondo regime as explained below.

There is a second Kondo regime that appears when $t''$ is increased. In this new regime two lateral peaks appear in the LDOS of the non interacting central site, together with an important reduction of the LDOS at $E_f$ (see Figure \ref{figura2} B). As explained above, this is because the energy broadening of the molecular states with weight in it is now lower than their energy separation $\Delta E \sim t''$. These molecular orbitals of higher and lower energy are empty and double occupied, respectively. Remembering that the central site has weight mostly on these levels, we can conclude that with the non interacting central site occupied with (zero) or two electrons no internal screening is now possible. The molecule has a net spin $S\approx 1/2$, as shown if Fig. \ref{figura2} B, inset, equally distributed between the two dots, which become Kondo correlated with the spins in the leads. The net spin is not exactly $1/2$ since the molecule is connected to the leads through $t'$. It can be checked that, decreasing $t'$, the value of the total spin tends to $1/2$ as $t''$ increases. Although the molecule is connected to four leads, it represents all the same a traditional bulk one channel SU(2) Kondo effect, as the electrons can flow freely from one channel to the other \cite{Potok07}. In Fig. \ref{figura3} we observe a three peak structure in the transmission for $t'' \gtrsim 0.04$, being the central one associated with a molecular Kondo effect. The lateral peaks correspond to tunneling resonances in which the molecule has a charge that fluctuates around $4$ (left peak) or $2$ (right peak) and the total spin is close to zero. This molecular Kondo regime represents a system that has a state at the Fermi level located mainly at the QDs and providing a channel for the electrons to go from one dot to the other. So, similarly to the previously analyzed two-Kondo temperature regime, the conductance through the upper left and upper right leads is $e^{2}/h$, half of its maximum value $2e^{2}/h$ reached when $t''=0$, as shown in Fig. \ref{figura4} (A).

In Figs. \ref{figura4} (B) and (C) we present, respectively, finite-U SBMFA results showing the renormalized energy level $\tilde{\epsilon}_{i}$ of the local state in the QDs and the renormalization parameter $\bar{Z}^2$ as a function of the gate potential $V_{g}$ applied on the QDs and for some of the values of the connections $t''$ used in Figs. \ref{figura2} and \ref{figura3}. For all values of $t''$ we observe a plateau structure in $\tilde{\epsilon}_{i}$  which, in the context of the finite-U SBMFA, is the fingerprint of the Kondo effect. This plateau is an indication of the existence of a resonance, the Kondo peak, in the LDOS of the QDs, that remains fixed near the Fermi level as the potential $V_{g}$ varies. This result, together with the renormalization observed in $\bar{Z}^2$, corroborates the Kondo nature of the two states described in this section. It is important to observe that the two-Kondo temperature regime, obtained for $t'' \lesssim 0.04$, is associated to a larger plateau structure in $\tilde{\epsilon}_{i}$ and to a stronger renormalization (lower $\bar{Z}^2$) of the connections if compared to the Kondo molecular regime. 

The magnetic RKKY interaction between impurities was carefully studied in a system similar to our's in Ref. \cite{Simon05}, taking into account the discrete character of the spectrum, with the conclusion that the equivalent to our $N=1$ (odd N in general) case ("at resonance'') implies a ferromagnetic correlation. Keeping in mind what was pointed out in the Introduction, that besides the RKKY interaction there can be other types of magnetic interactions in systems like these, we have evaluated the spin-spin correlations through a MCL calculation. The results show that the correlation between the dots and the central site is AF, while between them a ferromagnetic correlation is established. Taking into account the results presented in the inset of Fig. \ref{figura2} B, we can conclude that what determines the physics of our system is the fact that, as $t''$ increases, the two QDs plus the central site behave as a single entity, a molecule with $S \approx 1/2$ as we explained above, and this results in the molecular Kondo regime, which is reflected in the LDOS, transmission and conductance of the system.

\subsection{N=2}

\begin{figure}
\centering 
\rotatebox{0}{\scalebox{0.33}{\includegraphics{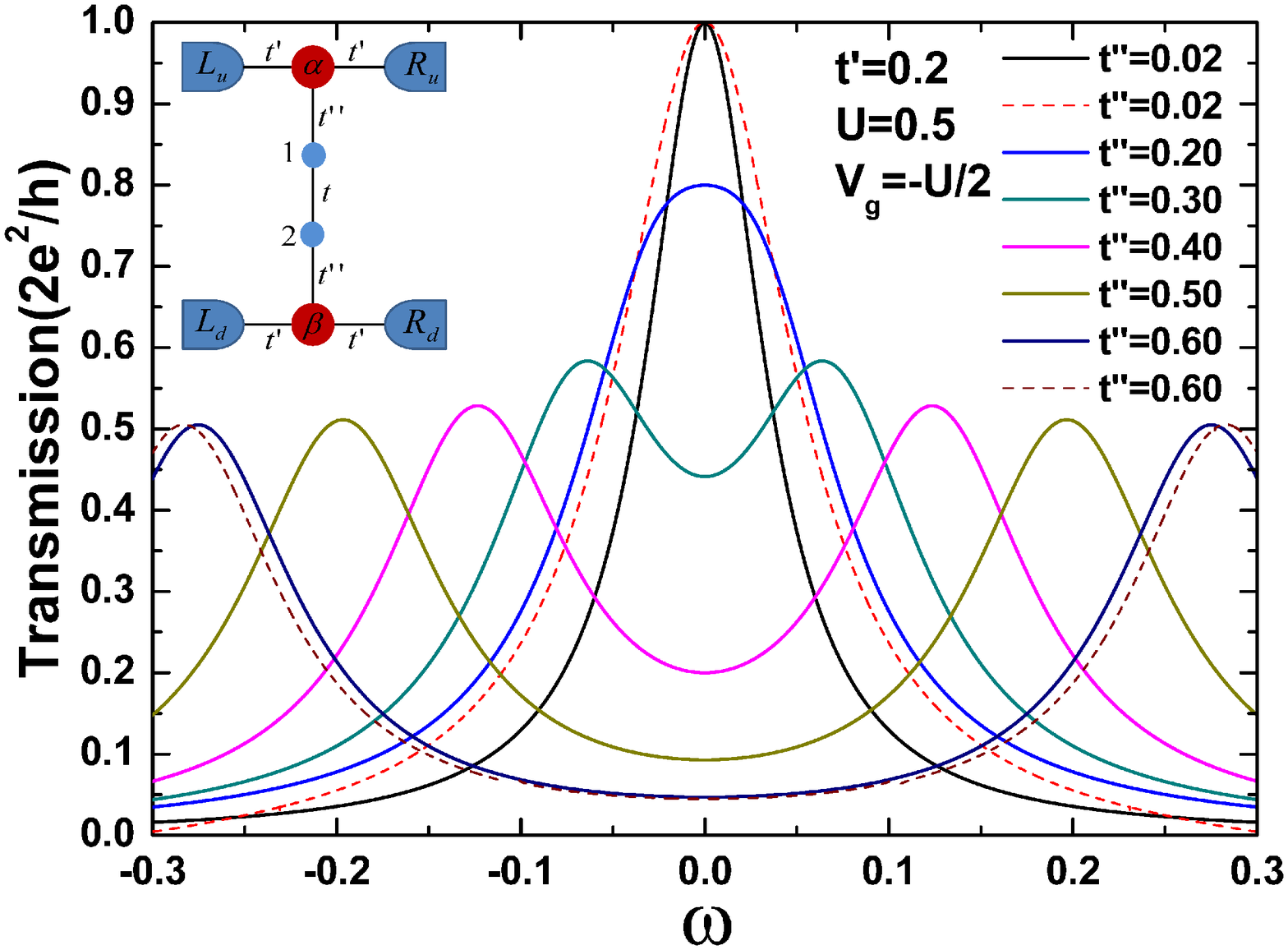}}}
\caption{(Color online) Transmission $T$ as a function of $\omega$ for different magnitudes of the connection $t''$ for the N=2 case  with $V_{g}=-U/2$. The continuous lines correspond the finite-U SBMFA results, while the dashed lines correspond to MCL results. As for the N=1 case, the agreement between both methods is very good.}
\label{figura5}
\end{figure}

\begin{figure}
\centering 
\rotatebox{0}{\scalebox{0.5}{\includegraphics{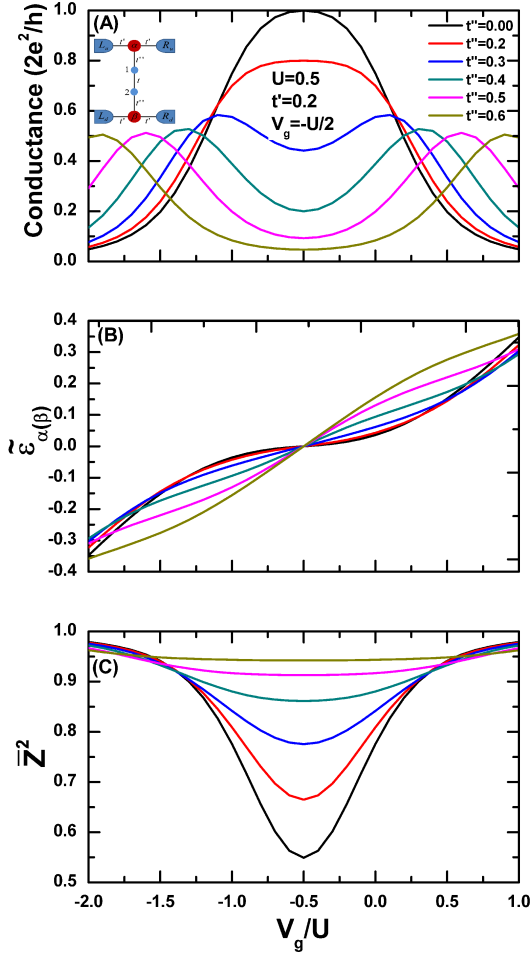}}}
\caption{(Color online) Conductance through the upper leads (A), renormalized energy level $\tilde{\epsilon}_{\alpha(\beta)}$ (B) and the parameter $\bar{Z}^2$ (C), as a function of the gate potential $V_{g}$ applied in the basis of the QDs for the system with different magnitudes of the connection $t''$ with the central NILC.} 
\label{figura6}
\end{figure}

The two dots system defined with $N=2$ corresponds to a four atoms molecule with two non-interacting intermediate sites. In this case the molecule has four levels, two well above and below the Fermi level, bonding and antibonding states created by the strong connection between the two sites forming a singlet. The other two levels are separated from the Fermi level by an amount proportional to $t''$. In Fig. \ref{figura5} we present the transmission $T$ between the leads left(up) and right(up) for the system with $V_{g}=-U/2$ and, therefore, for a charge occupation of four electrons in the molecule. Results for the LDOS, as for the N=1 case, are very similar to the transmission, and are not presented to avoid repetition. Again, the agreement between finite-U SBMFA and MCL is very good. The results do not present a dip in the LDOS of the QDs for small values of $t''$ and therefore it is not observed any signal of two coexisting Kondo temperatures, like in the $N=1$ case. Instead, a single Kondo peak is observed. This is so because no internal screening is possible as the two free electrons in the center occupy the antibonding state described before. Each QD has a spin which is only screened by the leads, giving rise to a Kondo effect at each QD characterized by the resonance observed in the transmission $T$ for $t''\leq0.3$. Consequently, the conductance, presented in Fig. \ref{figura6} (A) reaches it's maximum value $2e^{2}/h$.

Increasing $t''$ the system goes into a crossover regime characterized by the interplay between the Kondo ground state and the AF correlation between the spins of the QDs. These states are associated, respectively, to two energy scales: the Kondo temperature $T_{K}$ and the value of the effective AF interaction $I$. In Fig. \ref{figura5} we can identify the crossover region for $0.02<t''<0.3$ and observe, for $t''>0.3$, the splitting in the transmission $T$ (or in the LDOS, that has the same form as has been mentioned) due to the increasing of $I$. This splitting characterizes the establishment of the AF regime in the system \cite{Ribeiro12,Busser00}. We have calculated with MCL the spin-spin correlations (not shown) between each dot and the rest of the sites of the system, verifying that, as $t''$ increases, the (Kondo) AF correlation of each dot with the reservoirs to which it is directly connected decreases, while the also AF correlation between the two dots increases, giving support to the well established image of the Kondo-AF crossover mentioned. Hence, in this case, we can conclude that as $t''$ increases, the two QDs plus the two non-interacting central sites also form a molecule, but with net spin $S=0$.

In Fig. \ref{figura6} the conductance (A) and the slave bosons parameters $\tilde{\epsilon}_{i}$ (B) and $\bar{Z}^2$ (C) are presented as a function of $V_{g}$. We observe that with the transition from the Kondo to the AF regime the conductance goes to zero in $V_{g}=-U/2$. We also conclude from an inspection of Fig. \ref{figura6} (B) and (C) that $\bar{Z}{\mapsto} 1$ together with the destruction of the plateau in $\tilde{\epsilon}$, indicating, in the context of the finite-U SBMFA, that the system is driven out of the Kondo regime with the increasing of $t''$. Considering the dependence of the conductance on $V_{g}$, we observe in (A) the formation of two lateral peaks. These peaks correspond to the molecular Kondo resonances \cite{Ribeiro12, Anda08} associated to a charge occupation of approximately five (left peak) or three (right peak) electrons and a total spin close to $0.5$. In this regime, transmission across the upper channel has a maximum value of $e^{2}/h$, half of the quantum of conductance, because there is an inter-dot flow of electrons through the molecular state that is half populated. The coherence existing between both dots in this Kondo molecular regime provides a channel for the conduction. Through the analysis of Fig. \ref{figura6} (B), within the context of the finite-U SBMFA, we observe that, as $t''$ increases, $\tilde{\epsilon}_{i}$ tends to form a double plateau structure \cite{note2} in the regions of $V_g$ corresponding to a charge occupation of approximately three or five electrons, corroborating the molecular Kondo regime, while the suppresion of the plateau for four electrons certifies that the system does not have a Kondo groundstate. We can conclude that this system behaves in a similar way as a two dots structure with a direct connection between them \cite{Ribeiro12,Hamad13}.

\subsection{Large $N$}

\begin{figure}
\centering 
\rotatebox{0}{\scalebox{0.45}{\includegraphics{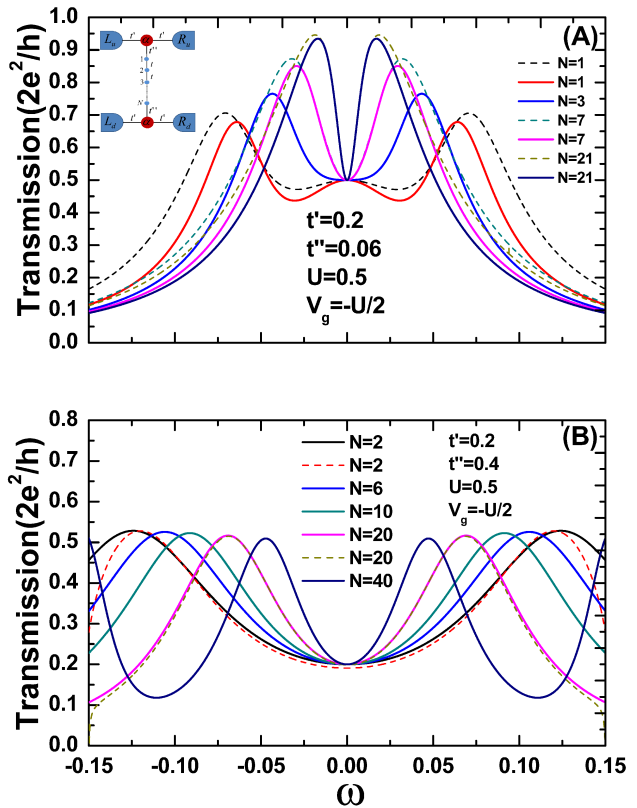}}}
\caption{(Color online) Transmission $T$ as a function of $\omega$ for different lengths of the NILC for the system with $V_{g}=-U/2$. The upper panel (A) corresponds to odd N with $t''=0.06$ while the lower panel (B) to the even N case with $t''=0.4$. The continuous lines correspond the finite-U SBMFA results while the dashed lines correspond to MCL results.}
\label{figura7}
\end{figure}

In Fig. \ref{figura7} we present the transmission $T$  between the leads $L_{u}$ and $R_{u}$ for the system with the gate potential adjusted to $V_{g}=-U/2$, and different lengths of the NILC. We present results for odd N with $t''=0.06$ and for even N with $t''=0.4$, since with these values of the connections the effects of varying N are most clearly seen, as the energies involved are different for each case. The LDOS in the QDs (not shown) behaves in the same way as the transmission. The description for large odd or even N is qualitatively similar to the cases $N=1$ and $N=2$ respectively. This occurs since N is low enough so as to preserve finite size effects, which happens if $\Delta \ll T_{K_{0}}$, being $T_{K_0}$ the isolated dot Kondo temperature. The behavior of the Kondo resonance alternates between even and odd N, as expected. 

In Fig. \ref{figura7} A, we clearly identify the existence of a regime of two coexisting Kondo temperatures for odd $N$ with the lowering of the width of the dip (related to the first Kondo temperature $T_{K_1}$) as N is increased. Since the central non-interacting sites are linked by a hopping $t>>t'' =0.06$, the level structure consists of one level at $E_f$, two Kondo peaks at $E_f$, which splitting is controlled by $t''$ and N, and $(N-1)/2$ levels above and below $E_f$. As in the case of $N=1$, $T_{K_1}$ is due to the existence of a NILC state at the Fermi level, which allows to localize two spins in the DQD and one free spin in the NILC. Hence, the two Kondo temperature regime develops in the same way as was described in the $N=1$ case. As N is increased, the weight of the state at the Fermi energy is spread along the odd sites of the NILC, reducing the weight of the local wavefunction, which is directly connected to the dots through $t''$. This effect reduces the splitting of the two Kondo peaks and hence the width of the dip decreases. Increasing N produces, in some extent, a similar result as reducing the effective value of $t"$.

For even N a similar phenomenon to the $N=2$ case takes place as N is increased since there is no energy level at $E_f$. The two lateral peaks that exist for a chosen value of $t''$ are pushed to energies nearer to the Fermi level, as can be seen in Fig \ref{figura7} B. Eventually, as N increases, additional peaks appear in the transmission close to the Fermi level, as can be observed in the $N=40$ case, where two additional lateral peaks are observed. A simple Green's function calculation, permits to verify the fact that the LDOS or the transmission T, at $\omega=0$ is independent of N, as long as N is low enough so as to keep finite size effects predominant. This is also observed in the mentioned Figure. Naturally, for a large number of intermediate sites $N$, the level separation in the NILC is comparable to the isolated-dot-Kondo-temperature and the system begins to behave as in the continuum limit \cite{Thimm99}, with a typical Kondo peak at LDOS of the dots.

\section{Concluding Remarks}
\label{Conclusions}

We have studied a system of two QDs connected with infinite leads and between themselves through a NILC. We have emphasized the quantum box character of this channel connecting the two dots. These kind of systems are very interesting since they can be used as quantum gates and they are experimentally feasible. Nowadays it is possible to manipulate QDs and even single atoms or molecules that act as magnetic impurities, and to control precisely the intermediate number of non-interacting sites through which the interaction between them is mediated \cite{Neel11}. For the particular geometry assumed, we analyzed the dependence of the transport properties and the LDOS of each QD as its connection to the NILC $t''$ is varied. We studied in detail the  $N=1$ and $N=2$ cases, representative of the odd and even N respectively, and also presented results for larger $N$. Our calculations were done using both the finite-U SBMFA and  MCL, which showed a remarkable qualitative and quantitative agreement, giving support to our results. 

For the $N=1$ case, we could identify two regimes. For small $t''$, there is a double Kondo temperature regime, one being representative of the bulk SU(2) Kondo regime of each dot spin screened by the spins of the leads to which it is connected, and the other being a Kondo box regime in which the screening is done by the spin of the electron occupying the level at the Fermi energy of the interdot chain. This is reflected in the LDOS and transmission as a peak with a dip at the Fermi level. As $t''$ increases, there is a crossover to a molecular Kondo regime where the two QDs plus the central site act as a whole entity with spin $S=1/2$, Kondo correlated with the conduction electron spins of the leads. The conductance across the upper leads reflects this behavior.

For the $N=2$ case, increasing $t"$ there is a crossover from a single impurity Kondo resonance to a splitted one that eventually disappears when  the antiferromagnetic correlations between the dots, mediated by the intermediate sites, is dominant. This crossover is reflected in the transmission through the upper leads.

For a NILC with larger $N$, the behavior obtained is similar to the $N=1$ (for odd $N$) and $N=2$ (for even $N$) cases. For odd number of sites in the NILC and low values of the connection $t''$ with the dots, a two Kondo temperature regime is developed. The width of the dip (related to one of the Kondo temperatures) is reduced as N increases. In the even N case, and for values of $t''$ big enough so as to have splitted peak in the LDOS or transmission, the splitting is reduced as N increases. In both cases the behavior is a consequence of the renormalization of the energy spectrum as N is increased.

In general it can be concluded that, for the type of systems we have studied in this work, it is the structure of energy levels in the molecule composed by the two dots plus the central site(s), together with the fact that the QDs are initially Kondo correlated with the leads to which they are connected, what determines the structure of the LDOS and the transport properties of the system. Varying the connection $t''$ of the dots with the NILC changes the energy levels and produces crossovers that are reflected in the LDOS and in the transport properties.  

\acknowledgments
We acknowledge financial support from the brazilian agencies FAPERJ (CNE) and CNPq, the spanish MCYT (grant FIS2009-10325 and FIS2012-35880), CONICET (Argentina), Universidad de Alicante, and from PUC-Rio de Janeiro. We acknowledge fruitful conversations with C. A. B\"usser, L. O. Manuel, C. J. Gazza, P. Roura-Bas.


\begin{thebibliography}{99}
\bibitem{note1} Although as $t''$ increases there is a reduction of the DOS at the Fermi Level in the central site, it remains always finite, which would allow a current to flow through it.
\bibitem{note2} The double plateau structure is more clearly visible as the connection to the leads $t'$ decreases.
\end{thebibliography}
\end{document}